\documentclass{ptapap}
\usepackage{amsmath}
\usepackage{amssymb}

\author{Agnieszka Janiuk}[CFT]
\author{Joseph Saji}[CFT]
\author{Gerardo Urrutia}[CFT]
\affil[CFT]{Center for Theoretical Physics, Polish Academy of Sciences, Al. Lotnikow 32/46, 02--668, Warsaw, Poland}

\title{What we can learn about compact binary mergers from their kilonova signals?}
\headtitle{Kilonovae from binary mergers}

\begin{document}

\maketitle

\begin{abstract}

Compact binary mergers are sources of gravitational waves, and can be accompanied by electromagnetic signals. We discuss the possible features in the kilonova emissions which may help distinguish the black hole - neutron star mergers from the binary neutron stars. In addition, the amount of ejected material may depend on whether the system undergoes the creation of a transient hyper-massive, differentially rotating neutron star. In this context, the numerical simulations of post-merger systems and their outflows are important for our understanding of the nature of short GRB progenitor systems.
In this article, we present a suite of GR MHD simulations performed by the CTP PAS astrophysics group, to model the neutrino driven disk winds and their contribution to the kilonova emissions. The contribution of the disk wind to the jet collimation and variability is also briefly discussed.

\end{abstract}

\section{Introduction}

Short Gamma Ray Bursts (GRBs) are transient events of high energy radiation 
released via pulses, lasting from miliseconds up to couple seconds. 
The central engine which powers these explosions is composed of a rotating Kerr black hole and a magnetized accretion disk, extremely hot and dense ($T>10^{9} K$, $\rho > 10^{8}$ g cm$^{-3}$).
Under such conditions, the electron gas is degenerate and beta reactions take place. The microphysics and equation of state (EOS) of the gas is governed by nuclear physics, and  the EOS is no longer described by an adiabatic law\footnote{this law stands an analytic fomula, with $p=(\gamma-1)u$, where $p$ denotes pressure, $u$ is internal energy, and $\gamma$ is an adiabatic index, of a value between $4/3$ and $5/3$}.
The EOS provides a closing equation for the system of magnetohydrodynamic (MHD) equations that describe accretion flow evolution, to capture the gas microphysics. 


Compact binaries coalescence (binary neutron star, NSNS, or black hole-neutron star, BHNS) have been modeled numerically 
since over 20 years, in the context of short GRB progenitors \citep{Shibata2000, Korobkin2012, Paschalidis2015}.
The postmerger system can be either a hypermassive neutron star, supported by its differential rotation, or a promptly formed black hole. Eventually, the relativistic jet is produced, and provides an electromagnetic counterpart to the gravitational wave (GW).
Depending on the observer's viewing angle, one may detect a short-weak GRB and its afterglow emission, from the shocked interstellar medium, as well as other counterparts, like a kilonova \citep{Metzger2019LRR}.


The GRBs prompt emission is accompanied by other electromagnetic observables, emerging on longer timescales. The discovery of a kilonova signal, associated with the GRB afterglow \citep{Tanvir2013}
has been predicted by \cite{LiPaczynski1998}. They proposed that the radioactivities from dynamical ejecta after the NSNS merger can power a distinct
electromagnetic signal, in Optical and Near-Infrared bands. Its characteristics is similar to the supernova, but is several orders of magnitude fainter: the absolute magnitudes are below $M_{V}=-16$ at the peak, and rapidly fading, by about $\sim 0.5$ mag per day.
These transients are quite rare, and their rate is estimated at below 1 per cent
of the core collapse supernova rate. 
However, after the pre-merger expansion of the dynamical ejecta, also the subsequent accretion can provide a contribution to the kilonova, if only the disk outflows are fast enough and not absorbed by precedent ejecta.

Such system can be modeled via numerical simulation \citep{2017PhRvL.119w1102S}, and the heavy unstable isotopes are formed in the rapid neutron capture process (r-process), while the ejected material is very
neutron rich.
The blue and red kilonovae components associated with GRB-GW170817 have been attributed to  post-merger dynamical ejecta, and accretion disk winds \citep{Kilpatrick}.

\section{Numerical code}

The phenomena such as the
jet launching via extraction of the black hole rotational energy, or the expansion of magnetised, neutrino-driven winds from accretion disk, are well described by ideal MHD simulations in general relativity (GR MHD).
The GR MHD code, HARM \citep{Noble_et_all_2006}, is a finite volume, shock capturing scheme that solves 
hyperbolic system of partial differential equations of GR MHD. 
The plasma energy-momentum tensor, $T^{\mu\nu}$, is contributed by gas and electromagnetic field: $T^{\mu\nu}={T_{\left(m\right)}}^{\mu\nu}+{T_{\left(em\right)}}^{\mu\nu}$, where

\begin{equation}
  \
{T_{\left(m\right)}}^{\mu\nu}= \rho h u^\mu u^\nu + p g^{\mu\nu};
\hspace{1cm}
{T_{\left(em\right)}}^{\mu\nu}=b^\kappa b_\kappa u^\mu u^\nu+\frac{1}{2} b^\kappa b_\kappa g^{\mu\nu} - b^\mu b^\nu.
\end{equation}
Here $u^{\mu}$ is the four-velocity of gas, $u$ denotes internal energy density,  $b^{\mu}$ is magnetic four-vector, and
$h$ is the fluid specific enthalpy.
The continuity and momentum conservation equations read:
\begin{equation}
\
(\rho u^{\mu})_{;\mu} = 0;
\hspace{1cm}
T^{\mu}_{\nu;\mu} = 0.
\end{equation}
They are brought in conservative form, by implementing the Harten, Lax, van Leer (HLL) solver to calculate numerically the corresponding fluxes. 
In addition, magnetic field is evolved according to Maxwell equations
\begin{equation}
^{*}F^{\mu \nu}_{;\mu} = 0,
\end{equation}
where $^{*}F$ is the Faraday tensor dual, and the ideal MHD conditions eliminate three degrees of freedom due to vanishing electric field. The zero divergence constraint for magnetic field is kept by the numerical scheme.

In the current work, we use the code HARM\_COOL\_EOS, recently developed in our group.
 The microphysics of dense nuclear matter in the GRB engine is treated with specific tabulated, composition-dependent EOS, with three parameters: rest-mass density, temperature, and electron fraction. 
 It has to be noted that in this case the inversion between the EOS ('primitive') parameters and the variables advanced by the MHD scheme ('conservative') is non-trivial task. The code needs to solve numerically five coupled non-linear equations through the recovery scheme. After testing several of such schemes, we implemented
 the bracketed root-finding 1D scheme proposed by \citet{Palenzuela2015}, which gives minimum errors for the broadest range of temperatures and densities, in comparison to other tested, 2D and 3D schemes \citep{Janiuk_Potor}.

\section{Results}

\begin{figure}
  \centering
  \begin{minipage}{0.45\textwidth}
    \includegraphics[width=\textwidth]{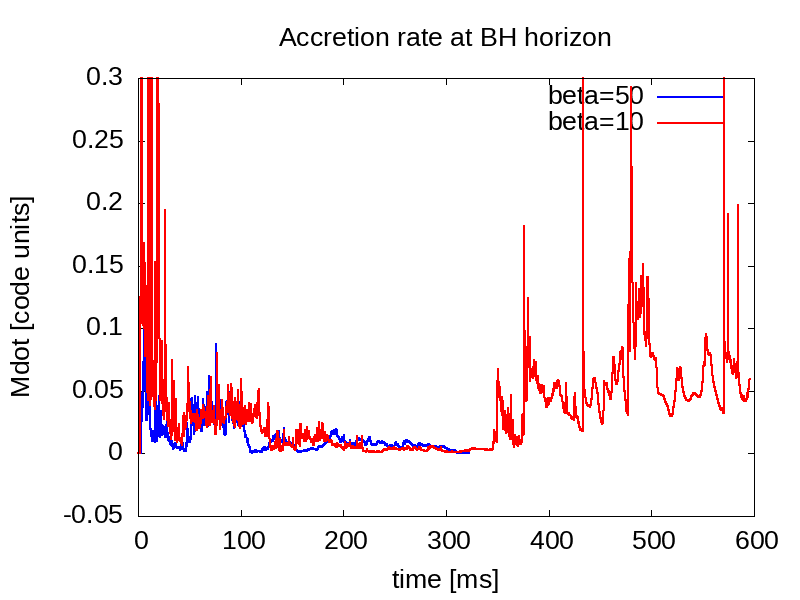}
    \caption{Accretion rate onto black hole as a function of time. Red line shows model with $\beta=10$, and thick line is for $\beta=50$.}
    \label{fig:mdot}
  \end{minipage}
  \quad
  \begin{minipage}{0.45\textwidth}
    \includegraphics[width=\textwidth]{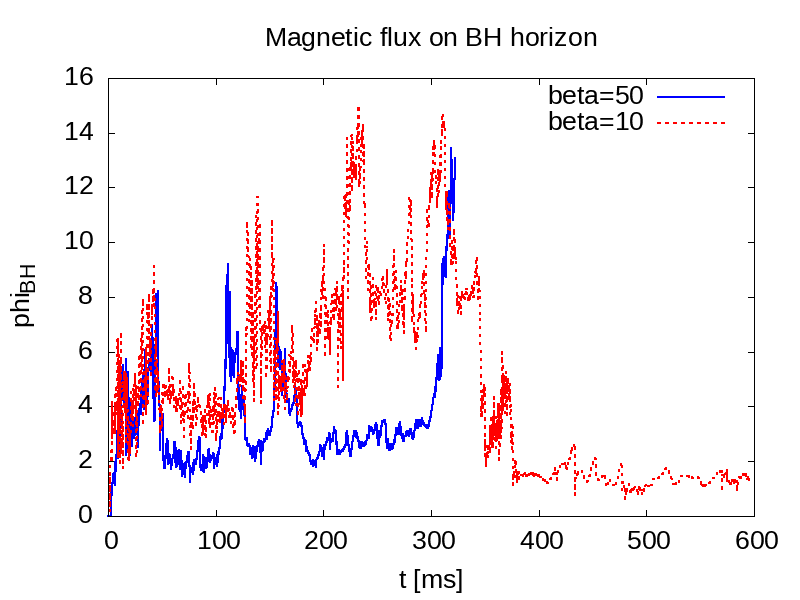}
    \caption{Normalized magnetic flux on the black hole horizon,
      as a function of time for the same models as in Fig. \ref{fig:mdot}}
    \label{fig:phiB}
  \end{minipage}
\end{figure}

In this section we present preliminary results of the new implementation of the neutrino leakage scheme in HARM\_COOL\_EOS, reading the Helmholtz tables \citep{fryxell2000}, where the EOS of dense matter is taken as $P(\rho,T, Y_{e})$ and $\epsilon (\rho, T, Y_{e})$, for wide range of densities and temperatures. The evolving electron fraction is giving an additional source term to the energy equation:
The neutrino leakage scheme is adapted as in \citet{Rosswog} and computes a grey optical depth estimate along the radial rays for electron neutrinos, anti-neutrinos, and heavy lepton 'nux'. 

The quantities plotted in Figures  \ref{fig:mdot} and \ref{fig:phiB}, are showing the accretion rate through the BH horizon, and magnetic flux on the horizon, respectively. The accretion rate is anti-correlated with the magnetic flux, indicating that the magnetic barrier acts on the plasma, pushing it out from the horizon at some episodes. They are identified as 'magnetically arrested state' (MAD) and are more prolonged in the simulation with higher initial magnetisation of the disk. 


\begin{figure}
  \centering
  \begin{minipage}{0.45\textwidth}
    \includegraphics[width=\textwidth]{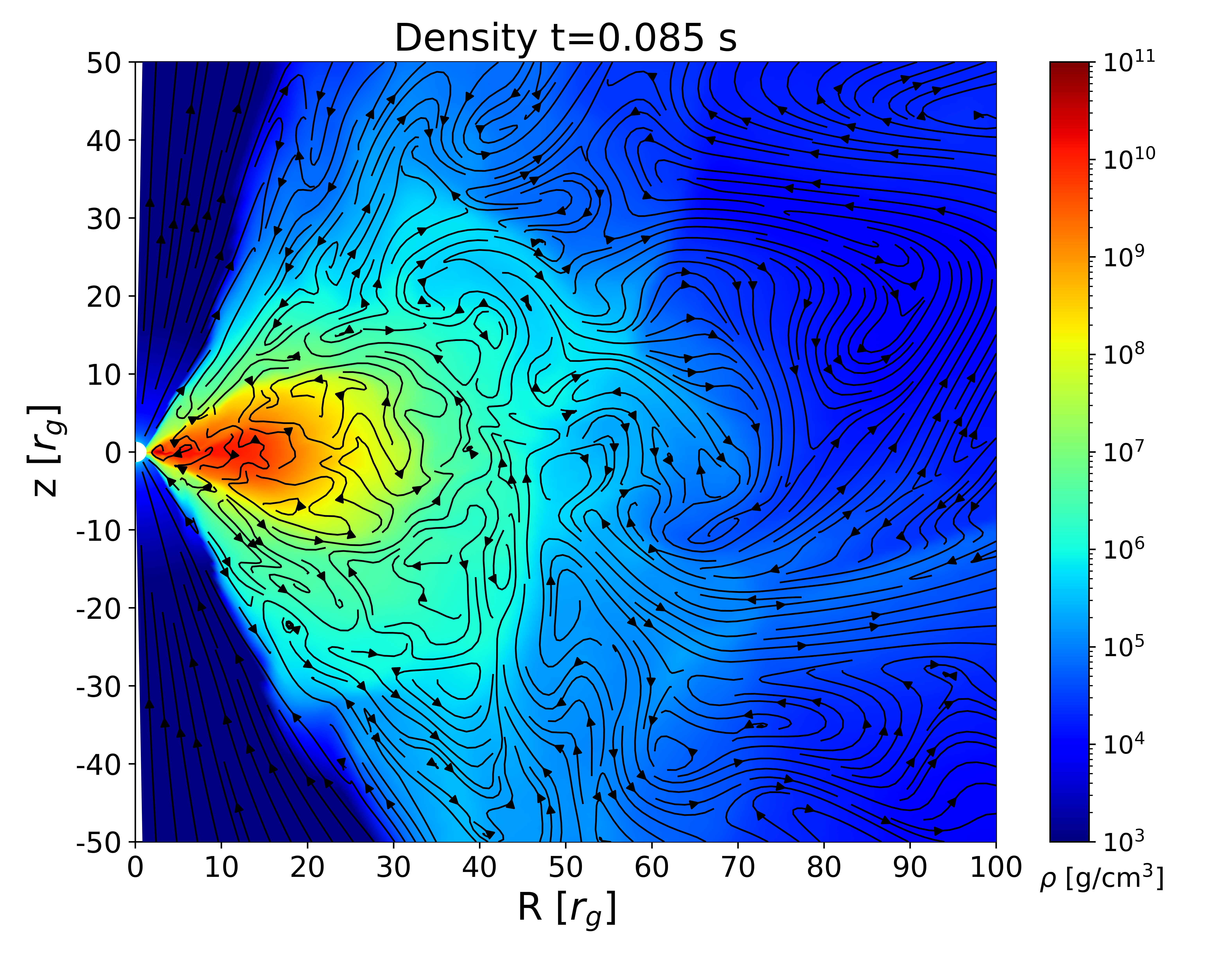}
    \caption{Density and magnetic field streamlines in the innermost regions of the central engine. Snapshot is taken at evolved time of $t=85$~ms, in the middle of the simulation.}
    \label{fig:density}
  \end{minipage}
  \quad
  \begin{minipage}{0.45\textwidth}
    \includegraphics[width=\textwidth]{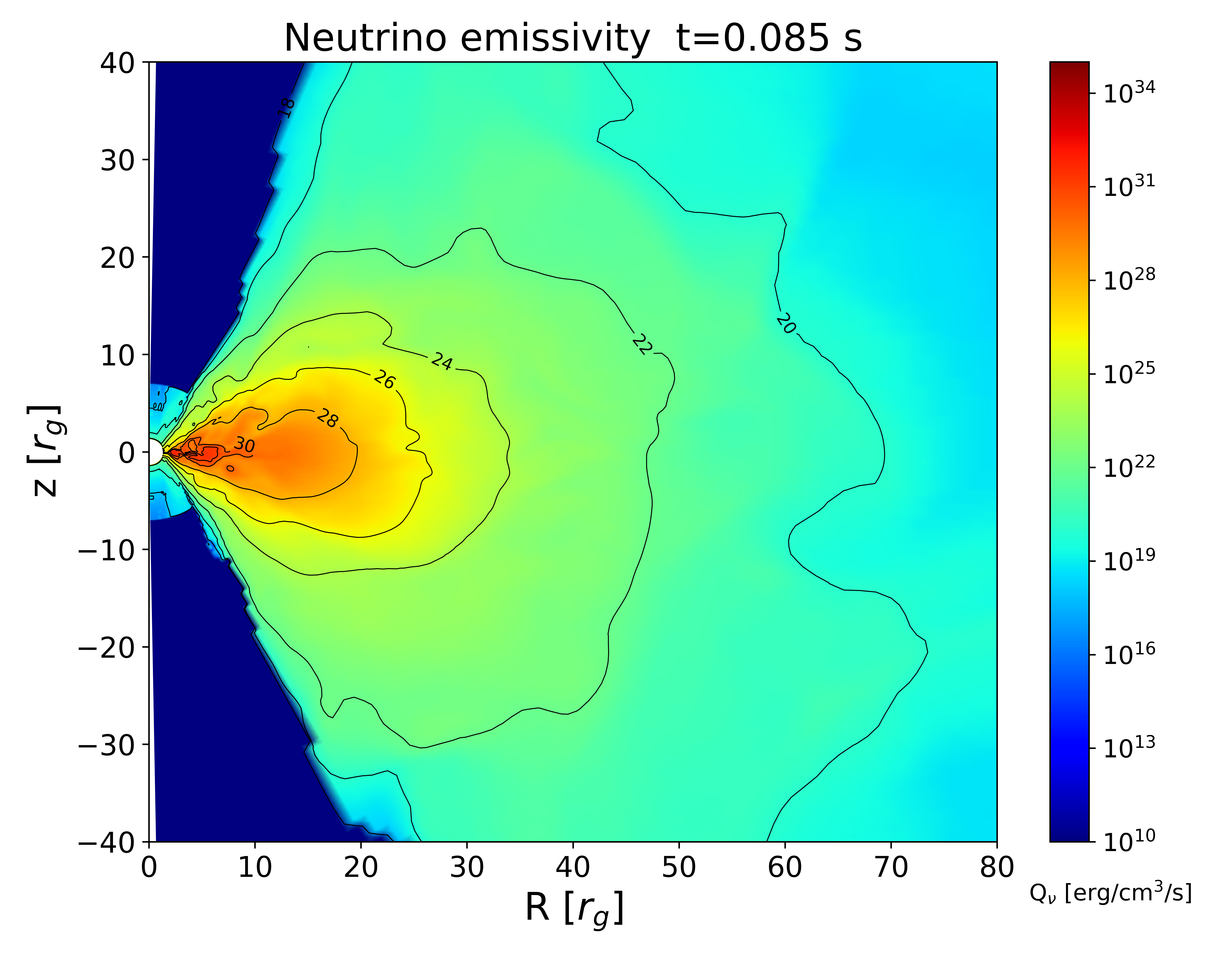}
    \caption{Neutrino emissivity map in the GRB central engine, for the same model and snapshot as in Fig. 1. Parameters are black hole mass of 3 $M_{\odot}$ and its spin $a=0.9375$}
    \label{fig:neutrino}
  \end{minipage}
\end{figure}

\begin{figure}
  \centering
  \begin{minipage}{0.45\textwidth}
    \includegraphics[width=\textwidth]{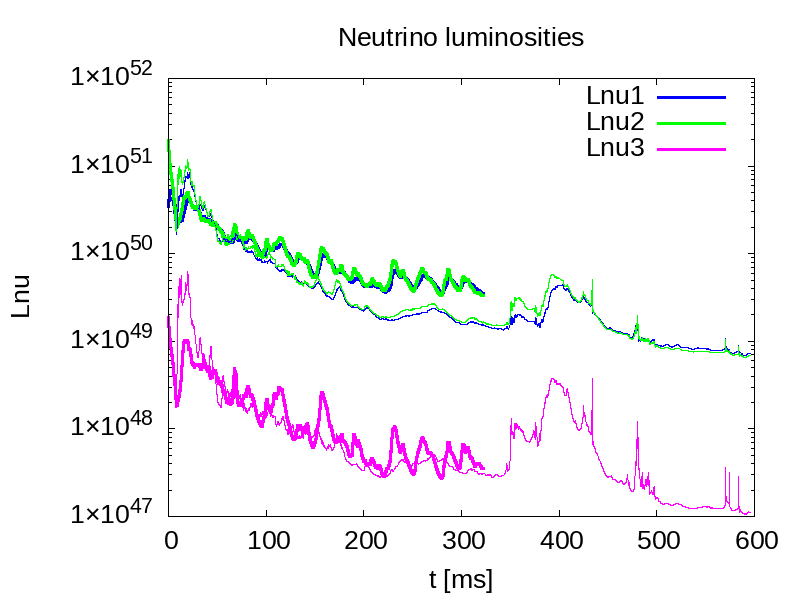}
    \caption{Neutrino luminosity as a function of time. Blue and green lines show electron neutrino and antineutrino fluxes (Lnu1 and Lnu2), integrated over volume in the disk. Purple lines show the heavy lepton neutrino fluxes (Lnu3). Thin lines show model with $\beta=10$, and thick lines are for $\beta=50$.}
    \label{fig:Lnu}
  \end{minipage}
  \quad
  \begin{minipage}{0.45\textwidth}
    \includegraphics[width=\textwidth]{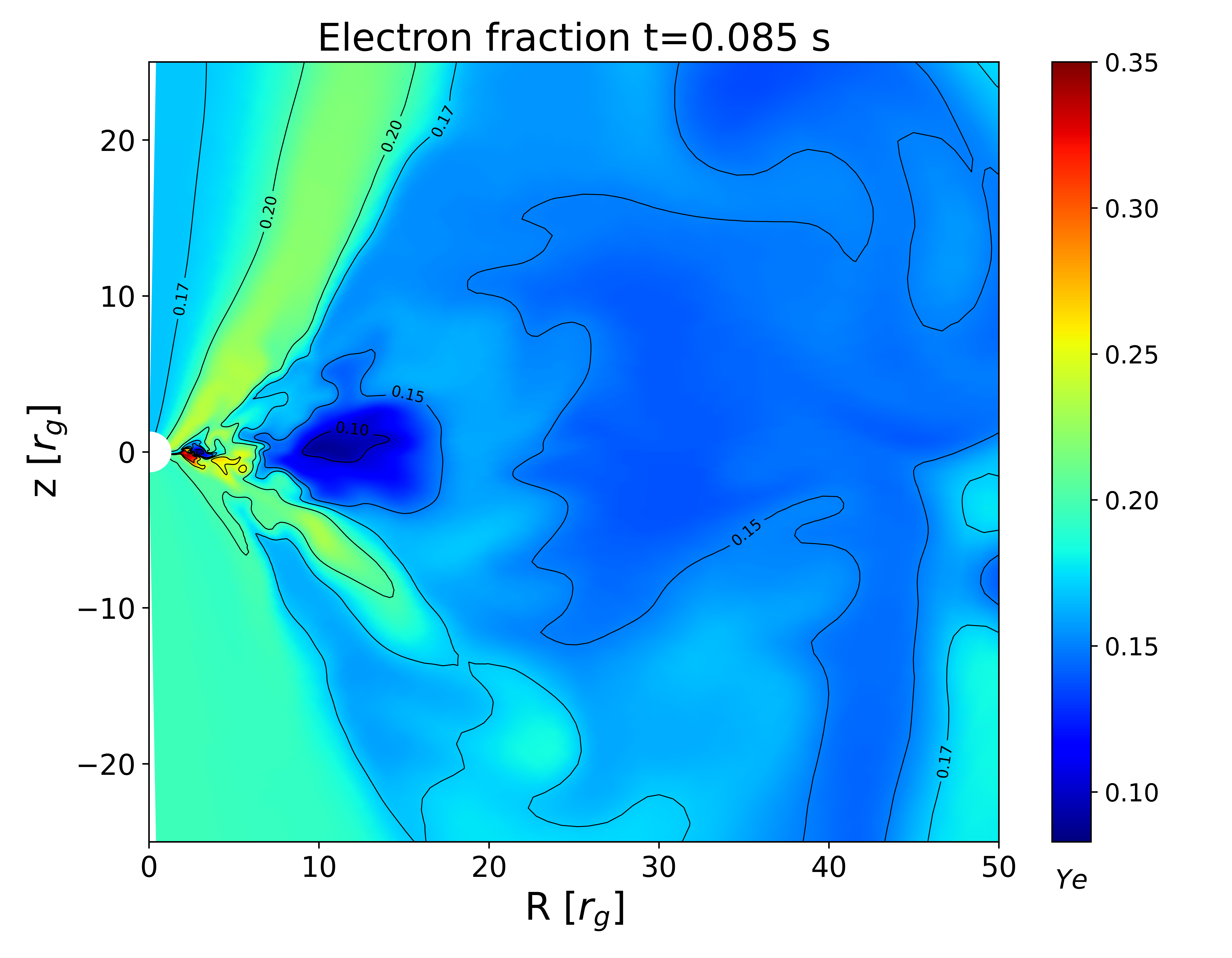}
    \caption{Electron fraction distribution map in the GRB central engine, for the same model and snapshot as in Fig. 1. Parameters are black hole mass of 3 $M_{\odot}$ and its spin $a=0.9375$, and initial gas-to-magnetic pressure ratio in the disk $\beta=50$.}
    \label{fig:Ye}
  \end{minipage}
\end{figure}

In Figures \ref{fig:density} and \ref{fig:neutrino} we plot the density distribution and neutrino emissivity for a chosen snapshot at the beginning of the simulation with $\beta=50$ (less magnetized initial condition, NSNS merger scenario). Turbulent accretion disk is located in the equatorial plane, and accretes in a 'standard and normal accretion episode' (SANE). Neutrino emissivity of the inner densest disk 
parts is as high as $10^{30}$ erg s$^{-1}$ cm$^{-3}$. The winds become unbound and driven by magnetic field only above a certain radius, $r\sim 30 r_{g}$.
The electron fraction distribution in disk and wind for the same snapshot is visualised in Fig. \ref{fig:Ye}.
\begin{table}
\centering
\begin{tabular}{|c||c|c|c|c||c|c|c|c|} 
 \hline
 Model & $r_{in}$ & $r_{max}$& $a$ & $S_{disk}$ & $M_{\rm BH}$ & $\beta$ & $Y_{e,0}^{\rm disk}$  & <$\dot{M}_{\rm out}$> \\  
 &[$r_{g}$]&[$r_{g}$]&&[s]&&&&[$M_\odot$~s$^{-1}$] \\
 \hline \hline

 FM50 & 4.0 & 9.0 & 0.9735 & 10.0 & 3 & 50 & 0.1  & $2.08\times 10^{-3}$\\  
 FM10 & 4.0 & 9.0 & 0.9735 & 10.0 & 3 & 10 & 0.1  & $2.57\times10^{-2}$\\ 

\hline

 \end{tabular}
    \caption{Model parameters of the neutrino cooled GRB engine.}
    \label{tab:params}
\end{table}
In Figure \ref{fig:Lnu} we plot the total (volume integrated) luminosity of neutrinos and antineutrinos, as a function of time for the whole simulation. Its shallow decrease with time only weakly reflects the changing mode of accretion (SANE vs. MAD), while oscillations in the energy output appear quasi-periodic.
The model parameters are summarized in Table \ref{tab:params}. We give here also the estimate for wind mass loss rates.

\subsection{Nucleosynthesis in GRB accretion disk winds}

Synthesis of elements heavier than Iron occurs via rapid neutron capture and conditions
 in the unbound outflows from the accretion disks in the GRB engine are plausible for that process.
 We have shown that the disk winds are an important source of mass ejection \citep{2017PhRvL.119w1102S, Janiuk2019}. The winds of a highly neutronized material, $Y_{e}\sim 0.1-0.4$, are driven by magnetic fields and neutrino heating, are expanding into the interstellar medium with mildly relativistic velocities, and entropy per baryon on the order of $12-15 k_{B}$.

The tabulated equation of state has been implemented in our code instead of the adiabatic law, and electron fraction evolution is driven by the lepton number conservation. 
By means of the tracer particles, we follow the nucleosynthesis on the streams expanding with the disk wind 
used as an input to the nuclear reaction network post-processing.
The network contains more than a thousand isotopes in its database, and takes into account the fission reactions and electron screening (see \citet{Lippuner2017} for details).
All three peaks of the Solar abundance pattern are reproduced. 
The second and third maximum are obtained for the Lanthanide- and Actinide-rich ejecta, 
with the electron fraction on the order of $Y_{e} \lesssim 0.1$. 

The detailed composition pattern of the outflow depends on the
engine parameters: the black hole spin and disk mass, as well as its 
initial magnetisation.
We postulate therefore, that different types of engines,
will imprint their properties in the observed kilonova signals due to various composition patterns.
\begin{figure}
  \centering
    \begin{minipage}{0.45\textwidth}
\includegraphics[width=0.8\textwidth]{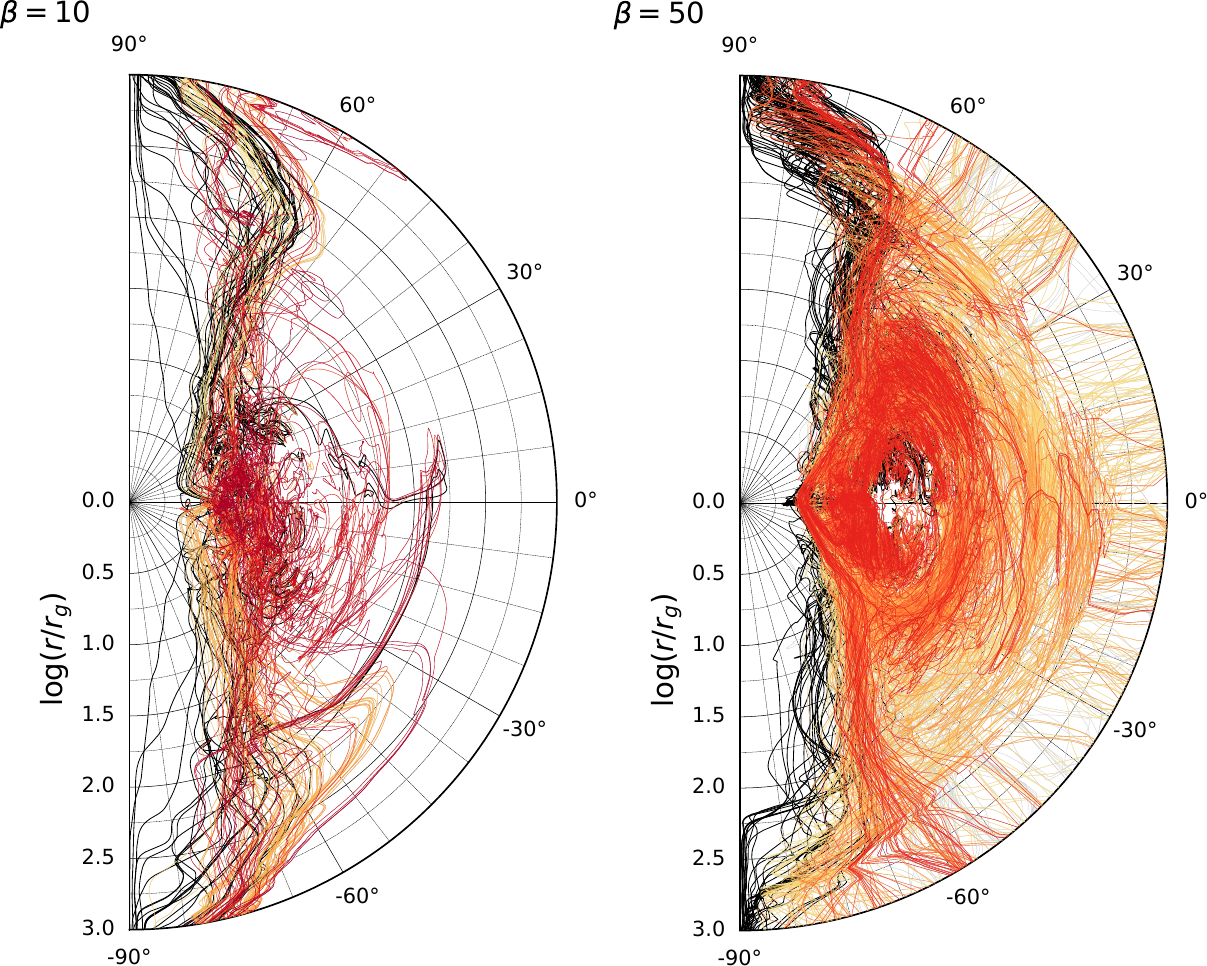}
    \end{minipage}
      \begin{minipage}{0.45\textwidth}
    \caption{Trajectories of wind particles for two models, representing engine of 
    binary neutron star merger. Black tracers represent jet outflows, orange-yellow ones are those
    representing unbound wind.Panles refer to model FM10 (left) and FM50 (right).}
    \label{fig:tracers}
    \end{minipage}
    \end{figure}
For instance, there are some distinguishable
features between BHNS and NSNS mergers. 
The average electron
fraction is higher for the BHNS case especially on the equator,
predicting a less effective r-process nucleosynthesis (Urrutia et al., in prep.)
The trajectories distributions for the two 
simulations listed in Table \ref{tab:params} are depicted in Fig. \ref{fig:tracers} and show that for FM50 postmerger system, the fastest tracers reach outer boundary at a later time, and larger polar angles, than for FM10 case.

\section{Discussion and Conclusions}

Our models do not provide direct  method to distinguish between BH-NS and NS-NS progenitors,
but still some features like the mass of ejected material and its composition, leading to different shape of kilonova lightcurves, are expected.
We can also speculate on the role of magnetic field strength, and possible MAD episodes, on the 
variability pattern of the neutrino luminosity. This would provide some quasi-periodic 
variable energy input to the jet,
and consequently further shape the wind and jet interface, and affect observed GRB emission 
(Urrutia et al. in prep; see also Urrutia \& Janiuk, this Proceeding). 
We address this issue to the GRB progenitor properties, and try verify the scenario in 
which the BHNS progenitors would rather produce less magnetised disks than NSNS systems (due to magnetic flux conservation).

On 11 December 2021, the Fermi GBM triggered and located GRB 211211A, also detected by the Swift/BAT. The GBM light curve consists of an exceptionally bright emission made up of three separate pulses with a duration  of about 34.3s (50-300 keV). 
It was not located in any star-forming region and associacion with supernova is excluded \citep{Troja2022}.
A kilonova component was nevertheless identified in the UV/optical/infrared by spectral analysis \citep{Rastinejad2022}.
This system seems a puzzle to our understanding of the short GRB progenitors, however according to \citet{Meng}
photosphere model with a structured jet can satisfactorily explain the peculiar long duration, through the duration stretching effect on the intrinsic longer ($\sim 3$ s) duration of NS-BH merger.
More evidence of the BHNS merger origin is found, especially the fit of the afterglow-subtracted optical-NIR light curves by the significant thermal cocoon emission and the sole thermal red kilonova component.

\acknowledgements{This research was supported in part by the grant 2019/35/B/ST9/04000 from Polish National Science Center. This research was carried out with the support of the Interdisciplinary Center for Mathematical and Computational Modeling at the University of Warsaw (ICM UW), and was also supported in part by PLGrid Infrastructure, under grant plg-grb6.
}

\bibliographystyle{ptapap}
\bibliography{janiuk_kilo}

\begin{thebibliography}{18}
\providecommand{\natexlab}[1]{#1}
\providecommand{\url}[1]{\texttt{#1}}
\providecommand{\urlprefix}{URL }
\providecommand{\eprint}[2][]{\url{#2}}

\bibitem[{{Fryxell} et~al.(2000)}]{fryxell2000}
{Fryxell}, B., et~al., \emph{\apjs} \textbf{131}, 1, 273 (2000)

\bibitem[{{Janiuk}(2019)}]{Janiuk2019}
{Janiuk}, A., \emph{\apj} \textbf{882}, 2, 163 (2019)

\bibitem[{{Janiuk}(2023)}]{Janiuk_Potor}
{Janiuk}, A., \emph{arXiv e-prints} arXiv:2303.18129 (2023)

\bibitem[{{Kilpatrick} et~al.(2017)}]{Kilpatrick}
{Kilpatrick}, C.~D., et~al., \emph{Science} \textbf{358}, 6370, 1583 (2017)

\bibitem[{{Korobkin} et~al.(2012){Korobkin}, {Rosswog}, {Arcones}, \&
  {Winteler}}]{Korobkin2012}
{Korobkin}, O., {Rosswog}, S., {Arcones}, A., {Winteler}, C., \emph{\mnras}
  \textbf{426}, 3, 1940 (2012)

\bibitem[{{Li} \& {Paczy{\'n}ski}(1998)}]{LiPaczynski1998}
{Li}, L.-X., {Paczy{\'n}ski}, B., \emph{\apjl} \textbf{507}, 1, L59 (1998)

\bibitem[{{Lippuner} \& {Roberts}(2017)}]{Lippuner2017}
{Lippuner}, J., {Roberts}, L.~F., \emph{\apjs} \textbf{233}, 2, 18 (2017)

\bibitem[{{Meng} et~al.(2023){Meng}, {Wang}, \& {Liu}}]{Meng}
{Meng}, Y.-Z., {Wang}, X.~I., {Liu}, Z.-K., \emph{arXiv e-prints}
  arXiv:2304.00893 (2023)

\bibitem[{{Metzger}(2019)}]{Metzger2019LRR}
{Metzger}, B.~D., \emph{Living Reviews in Relativity} \textbf{23}, 1, 1 (2019)

\bibitem[{{Noble} et~al.(2006){Noble}, {Gammie}, {McKinney}, \& {Del
  Zanna}}]{Noble_et_all_2006}
{Noble}, S.~C., {Gammie}, C.~F., {McKinney}, J.~C., {Del Zanna}, L.,
  \emph{\apj} \textbf{641}, 626 (2006)

\bibitem[{{Palenzuela} et~al.(2015)}]{Palenzuela2015}
{Palenzuela}, C., et~al., \emph{\prd} \textbf{92}, 4, 044045 (2015)

\bibitem[{{Paschalidis} et~al.(2015){Paschalidis}, {Ruiz}, \&
  {Shapiro}}]{Paschalidis2015}
{Paschalidis}, V., {Ruiz}, M., {Shapiro}, S.~L., \emph{\apjl} \textbf{806}, 1,
  L14 (2015)

\bibitem[{{Rastinejad} et~al.(2022)}]{Rastinejad2022}
{Rastinejad}, J.~C., et~al., \emph{\nat} \textbf{612}, 7939, 223 (2022)

\bibitem[{{Rosswog} \& {Liebend{\"o}rfer}(2003)}]{Rosswog}
{Rosswog}, S., {Liebend{\"o}rfer}, M., \emph{\mnras} \textbf{342}, 3, 673
  (2003)

\bibitem[{{Shibata} \& {Ury{\={u}}}(2000)}]{Shibata2000}
{Shibata}, M., {Ury{\={u}}}, K.~{\={o}}., \emph{\prd} \textbf{61}, 6, 064001
  (2000)

\bibitem[{{Siegel} \& {Metzger}(2017)}]{2017PhRvL.119w1102S}
{Siegel}, D.~M., {Metzger}, B.~D., \emph{\prl} \textbf{119}, 23, 231102 (2017)

\bibitem[{{Tanvir} et~al.(2013)}]{Tanvir2013}
{Tanvir}, N.~R., et~al., \emph{\nat} \textbf{500}, 7464, 547 (2013)

\bibitem[{{Troja} et~al.(2022)}]{Troja2022}
{Troja}, E., et~al., \emph{\nat} \textbf{612}, 7939, 228 (2022)

\end{thebibliography}

\end{document}